\documentclass[%
 reprint,
 amsmath,amssymb,
 aps,
]{revtex4-1}

\usepackage{graphicx}% Include figure files
\usepackage{dcolumn}% Align table columns on decimal point
\usepackage{bm}% bold math
\usepackage{epstopdf}

\newcommand{\zero}{|0\rangle}
\newcommand{\one}{|1\rangle}
\newcommand{\SLH}{(\mathbf{S},\mathbf{L},H)}

\newcommand{\beq}{\begin{equation}}
\newcommand{\eeq}{\end{equation}}

\begin{document}

%\preprint{}

\title{Photonic circuits for iterative decoding of a class of low-density parity-check codes}


\author{Dmitri S.~Pavlichin}
\email[Electronic address: ]{dmitrip@stanford.edu}
\author{Hideo Mabuchi}
\email[Electronic address: ]{hmabuchi@stanford.edu}
%\thanks{}
\affiliation{Edward L.\ Ginzton Laboratory, Stanford University, Stanford, CA 94305}


\date{\today}

\begin{abstract}
Photonic circuits in which stateful components are coupled via guided electromagnetic fields are natural candidates for native implementation of iterative stochastic algorithms based on propagation of information around a graph. Conversely, such message passing algorithms suggest novel circuit architectures for signal processing and computation that are well matched to nanophotonic device physics. Here we construct and analyze a quantum optical model of a photonic circuit for iterative decoding of a class of low-density parity-check (LDPC) codes called expander codes. Our circuit can be understood as an open quantum system whose autonomous dynamics map straightforwardly onto the subroutines of an LDPC decoding scheme, with several attractive features: it can operate in the ultra-low power regime of photonics in which quantum fluctuations become significant, is robust to noise and component imperfections, achieves comparable performance to known iterative algorithms for this class of codes, and provides an instructive example of how nanophotonic cavity quantum electrodynamic components can enable useful new information technology even if the solid-state qubits on which they are based are heavily dephased and cannot support large-scale entanglement.
\end{abstract}

\pacs{}
%\keywords{}

\maketitle

\section{Introduction}

Recent advances in the realization of nanoscale optical devices have shown the potential for ultra-low power integrated photonic circuits for classical information processing that would have significant advantages over electronic circuits in terms of heat generation and interconnect density~\cite{Beau11,Mill09}. In parallel, theoretical and computational tools have been developed for modeling the dynamics of photonic devices that have switching energies in the deeply sub-femtojoule, {\em few-photon} regime and are thus subject to quantum fluctuations~\cite{Kerc11}. These developments present an opportunity to consider the conventional (as opposed to quantum entanglement-enhanced) computational potential of such quantum noise-limited systems and to begin to consider architectural approaches that naturally accommodate noisy, low-power components interacting via coherent signal fields.

An intriguing source of architectural guidance is the broad and growing field of iterative, graph-based algorithms used today for computational tasks such as error-correction, probabilistic inference, optimization and signal processing~\cite{KollerFriedmanBook}. Such algorithms, including variants of message-passing schemes like belief propagation, have the flavor of nodes repeatedly exchanging information locally with their neighbors until global convergence. This picture invites an analogy to the dynamics of a network of photonic components, each of which has some internal degree of freedom (e.g., an `atomic' state), coupled via continuous interaction with propagating coherent fields. Thus photonic information processing systems could provide a native hardware platform for the implementation of iterative graph-based algorithms that are currently executed using electronic computers with incommensurate (though universal) circuit architectures that simulate message passing inefficiently.

Here we develop an instance of this direct mapping of a graph-based algorithm to a photonic circuit design for a simple and practically useful task: iterative decoding of {\it expander codes}, a class of low-density parity-check (LDPC) error-correcting codes for communication over a noisy channel. We work in the setting of linear coding theory in which every codeword is required to satisfy a set of parity check constraints, i.e., sums modulo 2 of subsets of its bits. The assignments (0 or 1) of the codeword bits and the values of their parity check sums correspond to the states ($\zero$ or $\one$) of a collection of two-state systems. Here we have in mind that $\zero$ and $\one$ ideally should correspond to orthogonal quantum states of an atom-like elementary physical degree of freedom, to facilitate ultra-low energy scales for switching, but our circuit does not require coherent superpositions or entanglement. For decoding a possibly corrupted channel output, we consider a simple iterative decoding procedure for the expander LDPC codes~\cite{SipserSpielman1994,SipserSpielman1996}: flip any bit (i.e., $0 \leftrightarrow 1$) that appears in more unsatisfied than satisfied parity check constraints; repeat until no more flips occur. We map this decoding procedure onto a closed-loop feedback circuit: a simple sub-circuit is engineered to encode parity check sum values in the state of an optical field, and another sub-circuit is designed to route feedback optical fields such that the states of certain components are flipped (i.e., $\zero \leftrightarrow \one$) at a rate that grows with the number of unsatisfied parity check constraints.

The proposed circuit is autonomous, continuous-time and asynchronous. No external controller, measurement system or clock signal is required, so the circuit can be realized as a single photonic device whose only required inputs are stationary coherent optical fields that drive the computational dynamics (i.e., supply power)~\footnote{Note that signal processing devices that require only optical forms of power may be of practical interest for large-area fiber optic networks.}. This follows the spirit of the systems we have designed in previous work on autonomous quantum memories~\cite{BitFlipCode,BSCode}. In contrast to our earlier work, the decoding circuit in the present proposal is straightforwardly extensible to the long block lengths (thousands of bits) used in practical LDPC implementations, as it involves a simpler feedback circuit architecture~\footnote{The circuits in our earlier work implement maximum likelihood (ML) decoders for small codes like the three-bit bit-flip code.  ML decoding is, however, impractical for larger block lengths, as it requires either a circuit size or decoding time exponentially large in the block length.  Iterative decoding algorithms such as the one discussed in this work have a larger error rate than ML decoders, but require only polynomial or even linear resources in the block length}.

Our circuit requires a collection of two-state latch systems coupled to input and output field modes. Here we consider designs based on the attojoule nanophotonic relay proposed in~\cite{Mabuchi2009}, which is based on ideas of cavity quantum electrodynamics (cavity QED), but any photonic system that functions as a latch potentially could be used in our circuit, e.g.,~\cite{Mabuchi2011}. Moreover, our scheme tolerates noisy components (e.g., spontaneous switching of a latch between the 0 and 1 states), can compensate for this noise with increased input optical power, and actually performs optimally (in terms of bits decoded per second) when the components ``misbehave" at some nonzero rate.  The graceful change in performance with increasing component imperfection and with varying optical input power is important for the practical usefulness of such a circuit. In our circuit design there is no real distinction between power and signal, as the power carried by the optical signal fields drives all the computational dynamics of the components, and it will be shown in Fig.~\ref{fig:VaryFeedbackProbePower} that simply increasing the optical input power reduces the error correction latency with fixed hardware. Our circuit tolerates a wide range of input powers with a constant performance as measured by bits corrected per joule.

This paper is organized as follows: We first briefly review linear error-correcting codes and an iterative decoding scheme for expander LDPC codes.  We then describe in an intuitive way the operating principles of our photonic circuit implementation of an iterative LDPC decoder. The subsequent section gives a more detailed picture of our circuit in terms of open quantum systems theory. We then present some numerical tests of our system and conclude with a discussion. The appendices describe circuit composition rules for open quantum systems, discuss the details of our numerical simulations, and derive some bounds for a parameter regime in which we expect our scheme to work.

\section{Linear Codes and Iterative Decoding\label{}}
We briefly review and set up notation for block binary linear error-correcting codes and an iterative decoding procedure for expander LDPC codes.

%%%%%%%%%%%%%%%%%%%%%%%%%%%%%%%%%%%%%%
\subsection{Linear Codes}

We work with binary bits transmitted in blocks of length $n$ through the binary symmetric channel (BSC) that with some fixed probability independently flips (i.e. $0\rightarrow 1$, $1 \rightarrow 0$) the transmitted bits.  To protect from errors, the sender restricts the possible channel inputs to the set of codewords---a subset of all $2^n$ possible inputs.  The decoder attempts to find the nearest codeword to the possibly corrupted output of the channel.  Equivalently, the bits are stored in memory that accumulates errors with time; the sender/decoder attempt to minimize losses through redundancy in the encoded memory bits.

%% define vector

Linear codes require each codeword $x^n = (x_1,\ldots,x_n)$ to satisfy $m$ parity check constraints.    A parity check constraint $\mathbf{c}$ is a subset of the $n$ message bits whose sum is constrained to equal $0$ modulo 2:
\beq
\sum_{j \in \mathbf{c}} x_{j} = 0 \pmod{2}
\eeq
A vector $x^n$ is a codeword if and only if it satisfies every constraint.  The rate $R$ of the code is the ratio of the number of non-redundant bits to the total number of bits per transmission, $R = (n-m)/n$.

It is useful to think of a code as an undirected bipartite graph, the Tanner graph \cite{Tanner1981}, whose $n$ `variable' nodes correspond to the message bits and whose $m$ `check' nodes correspond to the constraints.  Edges connect variable nodes and the constraints that include them.

%%%%%%%%%%%%%%%%%%%%%%%%%%%%%%%%%%%%%%
%\subsection{LDPC codes and expander codes}
%\subsection{ Expander codes $\subset$ LDPC codes $\subset$ linear codes}
\subsection{ Linear $\supset$ LDPC $\supset$ expander codes}

Low-density parity-check (LDPC) codes are linear codes introduced by Gallager in 1962 \cite{Gallager1962,Gallager1963} and are among the first known near capacity-achieving efficiently decodable codes.  The parity checks of a $(n,l,k)$ LDPC code all include $k$ bits, and each bit is included in $l$ parity checks (in the Tanner graph, each variable node has degree $l$ and each check node has degree $k$).   The codes are ``low-density'' because the total number of variable-check pairs is $l n$, linear in the block length $n$ (rather than quadratic in $n$ for a dense graph); the Tanner graph is sparse.  The rate of the code is $R=(n-m)/n = (k - l)/k$.

\begin{figure}
\includegraphics[width=8.6cm]{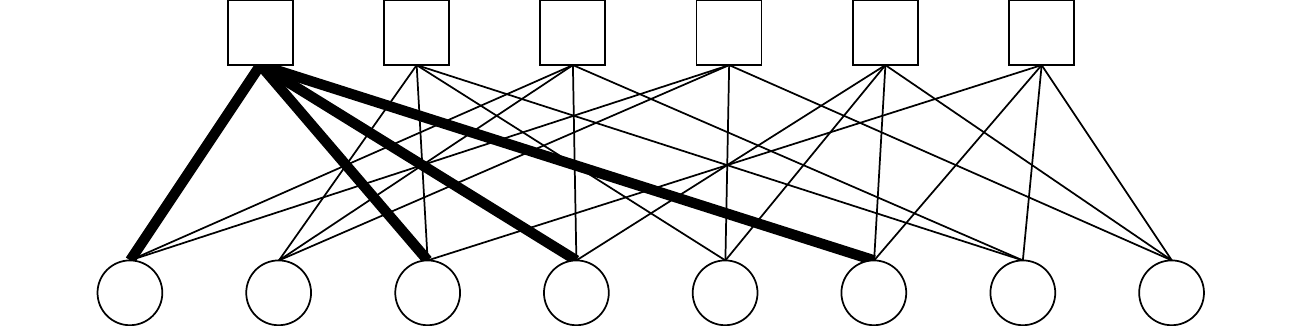}%
\caption{Tanner graph for a $(n= 8, l = 3, k = 4)$ LDPC code.  Circles (squares) indicate variable (check) nodes.  The thick edges indicate that parity check constraint 1 is $x_1 + x_3 + x_4 + x_6 = 0 \pmod{2}$ (numbering the check and variable nodes from left to right).}\label{fig:TannerGraph}
\end{figure}

Fig. \ref{fig:TannerGraph} shows the Tanner graph for a particular $(n= 8, l = 3, k = 4)$ LDPC code, where variable (check) nodes are drawn as circles (squares), and we have highlighted a particular parity check constraint.  This graph would look sparse for larger $n$.

%% other LDPC photonics suggestions

LDPC codes shine because they can be decoded efficiently by iterative algorithms that have good performance in practice and in theory.  These schemes include those in Gallager's original work \cite{Gallager1963}, as well as message-passing algorithms and belief propagation; for a theoretical analysis of their performance see \cite{LubyMitzenmacherShokrollahi2001,MackayNeal1997,Mackay1999,Richardsonurbanke2001}.  These schemes all have the flavor of variable and check nodes repeatedly exchanging information about the most likely codeword given the observed channel output and differ from each other in how that information is represented (e.g. binary or real-valued messages) and how new messages are computed from old.

Expander codes are a class of LDPC codes, introduced by Sipser and Spielman \cite{SipserSpielman1994,SipserSpielman1996}, for which a particularly simple iterative decoding procedure exists and which are easy to make by using a random construction.  Expander codes require the Tanner graph to be a good expander graph, meaning that the number of check nodes neighboring any small enough subset $V$ of the variable nodes grows fast enough linearly with $|V|$. For our purposes it suffices to note that a randomly sampled bipartite graph with fixed variable and check node degree (a regular LDPC code) probably makes a good expander code \cite{SipserSpielman1996}.

%%%%%%%%%%%%%%%%%%%%%%%%%%%%%%%%%%%%%%
\subsection{Iterative decoding of expander codes}\label{sec:iterativedecodingofexpandercodes}

The iterative decoding procedure that is our focus in this work is the sequential decoder of Sipser and Spielman \cite{SipserSpielman1996}. The variable bits are initially assigned to $0$ or $1$, equal to the observed output of the channel (we work with a binary symmetric channel that flips incoming bits with probability less than $1/2$).  The initial assignment of the variables may fail to satisfy all parity check constraints due to errors. The decoding procedure is as follows:

\begin{itemize}
\item Flip (i.e. $0\leftrightarrow 1$) any variable that is included in more unsatisfied than satisfied constraints.
\item Repeat until no more variables are flipped.
\end{itemize}

Each iteration reduces the total number of unsatisfied constraints, so the procedure terminates when either there are 0 unsatisfied constraints (successfully outputting a codeword) or it gets stuck and declares failure to decode.  While this procedure could be applied to any binary linear code, \cite{SipserSpielman1996} prove that for expander codes this procedure removes a constant fraction of errors and, if the initial fraction of errors is low enough, is guaranteed to succeed.  For the expander LDPC codes, each variable participates in $k$ constraints, so we flip the variable's assignment if the number of unsatisfied constraints is greater than $k/2$.

Importantly for our work, in \cite{SipserSpielman1996}'s numerical experiments, it was found that permitting the algorithm to make some amount of backwards progress (sometimes increasing the total number of unsatisfied constraints) increased the probability of success.  This suggests the procedure is robust to noise affecting the computation.  In our approximate implementation of this iterative algorithm, described below, backwards progress is unavoidable and the hardware itself is noisy, so this robustness of the decoding procedure to noise is desirable.

This procedure is not technically a message-passing algorithm in the sense of~\cite{Burshtein2001}, in that information flow from a check to a variable node (a possible ``flip" instruction) does not exclude information received by the check node from that variable node (the bit state). Nonetheless it is convenient to discuss the error-correcting dynamics, as~\cite{SipserSpielman1996} do, in terms of variable nodes receiving ``flip messages'' from check nodes.

%% photonic decoding circuit
%%%%%%%%%%%%%%%%%%%%%%%%%%%%%%%%%%%%%%
\section{A photonic decoding circuit: Overview}

We give an intuitive description of the operation of our expander code decoder circuit before giving a more precise description in terms of open quantum systems in the Section that follows.

%%%%%%%%%%%%%%%%%%%%%%%%%%%%%%%%%%%%%%
\subsection{The idea}

Our circuit consists of a collection of two-state ($|0\rangle$ or $|1\rangle$) systems, one for each of $n$ variable and $m$ check nodes in the Tanner graph for an error-correcting code.  Information exchange between the variable and check systems is mediated by coherent fields interacting with these systems (e.g. a beam scattering from one atom-cavity system into another).  There are two crucial interactions:

\begin{itemize}
\item Fields outgoing from a system can encode that system's state (perform a measurement)
\item Fields incoming to a system can drive that system into a desired state (apply a control)
\end{itemize}

\noindent These two interactions allow us to construct a closed-loop, autonomous measurement and feedback circuit that achieves:
\begin{itemize}
\item \textbf{Parity checks/Measurements}: A field scattered (e.g. a beam reflected) from the set of all variable bit systems included in some parity check constraint encodes their sum modulo 2.  This field then drives the check system into the $|\text{satisfied}\rangle$ or $|\text{unsatisfied}\rangle$ states ($|0\rangle$ or $|1\rangle$, respectively).

\item \textbf{Error correction/Feedback}: A field scattered from the set of all check systems that include a particular variable has an amplitude that increases with the number of unsatisfied checks involving that variable.  This field then drives the variable system to flip between the $|0\rangle$ and $|1\rangle$ state at a rate proportional to the magnitude of the field amplitude.  The more unsatisfied parity checks, the faster the flipping occurs.
\end{itemize}

The time evolution of this circuit is modeled as a continuous time Markov jump process~\footnote{This description follows from the open quantum systems dynamics treatment of our circuit.  See Sections \ref{sec:trajectories} for details}.  The jumps are changes in state ($\zero \leftrightarrow \one$) and the jump rates depend on amplitudes of fields interacting with the two-state systems.  The circuit is autonomous and asynchronous in that there is no external clock signal or external controller to process the parity measurement outcomes and to create an appropriate feedback field.

We note that the iterative decoding algorithm of \cite{SipserSpielman1996} that our circuit emulates, summarized in Section \ref{sec:iterativedecodingofexpandercodes}, can be cast in terms of a continuous time Markov jump process as well: if a variable is included in more unsatisfied than satisfied constraints, set the rate for ``flipping" it to $R_\text{flip} >0$, otherwise set $R_\text{flip}=0$.  In our implementation, the value of $R_\text{flip}$ scales with the number of unsatisfied constraints in a different way (and is never $0$; see Section \ref{sec:FeedbackToVariables}), but we attain comparable empirical performance in simulation.

Finally, we note that our circuit is essentially classical in its operation, even though we utilize quantum stochastic differential equations (QSDEs) to describe the dynamics of the components and their interactions in order to obtain a circuit model that is valid in the ultra-low power regime of significant quantum fluctuations (photon shot noise). Entanglement between different subsystems is insignificant and is not exploited, and thus does not need to be protected from interactions with the outside environment.

%%%%%%%%%%%%%%%%%%%%%%%%%%%%%%%%%%%%%%
\section{A photonic decoding circuit - construction}

We briefly review open quantum systems connected into circuits, describe the photonic component subsystems that make up our circuit, and specify their interconnection to form our iterative decoder circuit.  We give an intuitive description of our circuit's dynamics and defer a more detailed description to Appendices \ref{app:GoughJamesCircuitAlgebra} and \ref{app:Components}.

%%%%%%%%%%%%%%%%%%%%%%%%%%%%%%%%%%%%%%
\subsection{Open quantum systems and circuits}

We work in the framework developed by Gough and James \cite{GoughJames2008,GoughJames2009} for modeling open quantum systems interacting via coherent fields \cite{HudsonParthasarathy1984, Carmichael993, Gardiner1993, Barchielli2006}.  The basic component model (shown in Fig.~\ref{fig:SLH} of Appendix \ref{app:OpenQuantumSystems}) comprises a system with internal degrees of freedom coupled to incoming and outgoing field modes. The system is parametrized by its Hamiltonian $H$, by the coupling of the external modes to the internal degrees of freedom ($n$ by $1$ operator-valued vector $\mathbf{L}$), and by the way the incoming external field modes scatter into outgoing external field modes ($n$ by $n$ operator-valued unitary matrix $\mathbf{S}$).  The density matrix $\rho$ for the system's internal degrees of freedom evolves in time according to the master equation:

\beq \label{eq:mastereq}
\dot{\rho}_t = -i  [H,\rho_t] + \sum_{i=1}^n \left(L_i \rho_t L_i^\dagger - \frac{1}{2}\{L_i^\dagger L_i ,\rho_t\}\right)
\eeq

\noindent where $L_i$ is the $i$-th component of the external field mode coupling vector $\mathbf{L}$.  See Appendix \ref{app:GoughJamesCircuitAlgebra} for a more detailed discussion.

The Gough-James circuit algebra allows us to compute new $\SLH$ triplets in terms of old for two systems connected in series, in parallel, or for one system self-connected through feedback.  These composition rules are given in Appendix \ref{app:Circuits}.  A systematic, automated approach for specifying and simulating such circuits in software is presented in \cite{QHDL, ModelicaQuantumCircuitModels}.

%%%%%%%%%%%%%%%%%%%%%%%%%%%%%%%%%%%%%%
\subsection{Photonic circuit components} \label{sec:Components}

\begin{figure}[]%Relay%%%%%%%%%%%%%%%%%%%%%%%%%%%%%%%%
\includegraphics[width=8.6cm]{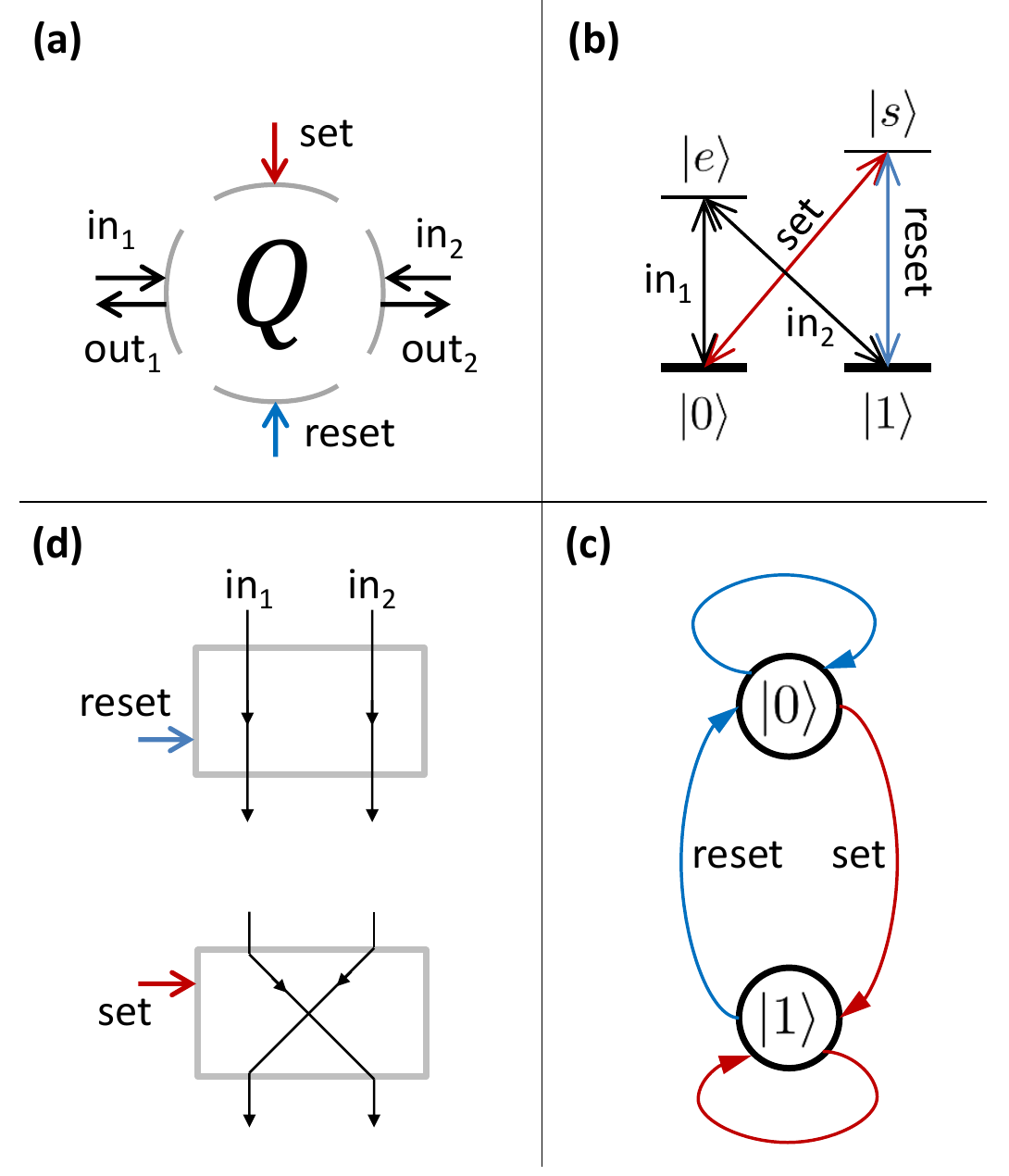}
\caption{Latch component from \cite{Mabuchi2009}.  \textbf{(a)} Input-output connections (reflected set and reset outputs not shown). \textbf{(b)} Input field couplings to internal states. \textbf{(c)} The latch approximated as a two-state continuous time Markov jump process after adiabatically eliminating the excited states $|e\rangle$ and $|s\rangle$ (see \cite{Mabuchi2009} for this derivation).  \textbf{(d)} The latch routes the input fields into output fields, switching them if its internal state is driven to $\one$.}\label{fig:Relay}
\end{figure}%%%%%%%%%%%%%%%%%%%%%%%%%%%%%%%%

The basic component of our circuit - used to represent both variable and check node assignments ($\zero$ and $\one$) - is a photonic latch, shown in  Fig. \ref{fig:Relay}, that behaves like the set-reset latch in electronics.  There are several proposals for implementing latching behavior in nanophotonic circuits \cite{Mabuchi2009, Mabuchi2011,VuckovicSwitch2012,VuckovicSwitch2008,NielsenKerckhoffSwitch}.  One such system, a coupled atom-cavity system \cite{Mabuchi2009}, is shown in Fig. \ref{fig:Relay} (panel (b)).  Our circuit construction is defined without reference to a particular physical system and assumes that the latch system that is used implements the following protocol.

The latch has a discrete internal degree of freedom (e.g. an atomic state) coupled to two external field modes, labeled ``set" and ``reset."  A signal incoming to the ``set" (``reset") input drives the latch into the $\one$ ($\zero$) state.  When neither the set nor reset input is powered, the latch maintains its current state.  Usefully for us, driving both the set and reset inputs simultaneously - an undefined condition for the electronic set-reset latch - results in astable behavior, with the latch state repeatedly jumping between the $\zero$ and the $\one$ state with exponentially-distributed jump times.

The latch routes two input channels (in$_1$ and in$_2$) into two output channels (out$_1$ and out$_2$).  When the latch is in the $\zero$ state, the outputs match the inputs ($\text{out}_{1,2} = \text{in}_{1,2}$); when the latch is in the $\one$ state, the outputs are switched ($\text{out}_{1,2} = \text{in}_{2,1}$).

In addition to the latch, our circuit uses beamsplitters with some fixed transmission and reflection coefficient.  Proposals for integrated nanophotonic beamsplitting devices include \cite{BayindirTemelkuranOzbay2000,Tao_etal2005}.  The Gough-James $\SLH$ description of these components connected to each other and driven by coherent fields is provided in Appendix \ref{app:Components}.

%%%%%%%%%%%%%%%%%%%%%%%%%%%%%%%%%%%%%%
\subsection{Circuit construction} \label{sec:CircuitConstruction}

We describe how the latches, beamsplitters, and coherent inputs are used to form our expander code decoding circuit.  There are two kinds of interactions to implement between the variable and check systems: parity check sums and feedback to ``flip" the variable nodes.

%%%%%%%%%%%%%%%%%%%%%%%%%%%%%%%%%%%%%%
\subsubsection{Parity checks} \label{sec:ParityChecks}

\begin{figure}[!t]%Parity check%%%%%%%%%%%%%%%%%%%%%%%%%%%%%%%%
\includegraphics[width=8.6cm]{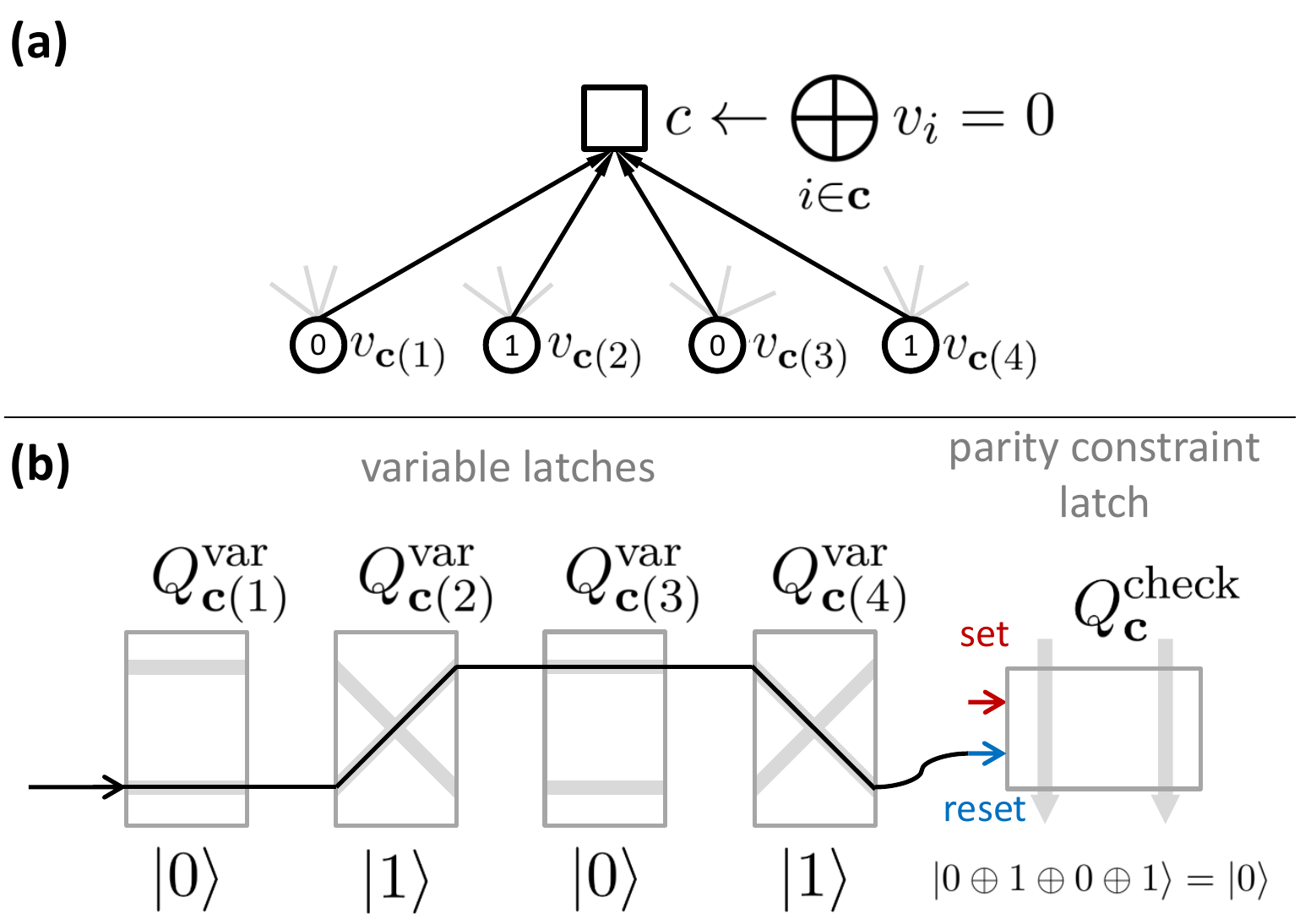}
\caption{Photonic circuit implementing a parity check computation.  \textbf{(a)} The parity check sum operation for a fragment of the Tanner graph for a linear code (rest of graph in gray).  The check node $\mathbf{c}$ is assigned the sum modulo 2 of the variable nodes.  \textbf{(b)} The photonic circuit implementation of the check sum using variable latches $Q_{\mathbf{c}(1)}^{\text{var}},\ldots,Q_{\mathbf{c}(k)}^{\text{var}}$ and check latch $Q^{\text{check}}_{\mathbf{c}}$.  Power is routed into either the SET or RESET ports of the check latch conditional on the parity of the variable latches' states.}\label{fig:ParityCheck}
\end{figure}%%%%%%%%%%%%%%%%%%%%%%%%%%%%%%%%

Fig. \ref{fig:ParityCheck} shows our parity check sum construction.  For each parity check $\mathbf{c}$ corresponding to the $k$-variable constraint $\bigoplus_{i=1}^k x_{\mathbf{c}(i)} = 0$, there are $k$ variable latch systems, $Q^\text{var}_{\mathbf{c}(1)},\ldots,Q^\text{var}_{\mathbf{c}(k)}$, and one check latch system $Q^\text{check}_\mathbf{c}$ (here $\oplus$ denotes addition modulo $2$).  The current assignment ($0$ or $1$) of the variables included in $\mathbf{c}$ is represented by the states ($\zero$ or $\one$) of the variable latches; the check latch's state is meant to represent the sum of these assignments modulo 2.  As shown in Fig. \ref{fig:ParityCheck}(b), the variable latches share two common optical paths for their in$_1$ and in$_2$ inputs and outputs.  An input field with amplitude $\alpha$ is incident to input port $\text{in}_1$ of check latch $Q_{\mathbf{c}(1)}^{\text{var}}$.  Subsequently, the two output ports of $Q_{\mathbf{c}(i)}^{\text{var}}$ connect to the two input ports of $Q_{\mathbf{c}(i+1)}^{\text{var}}$ for $i<k$.  The outputs of the final variable latch $Q_{\mathbf{c}(k)}^{\text{var}}$ connect to the set and reset ports of check latch $Q^{\text{check}}_\mathbf{c}$.

Each time a $\one$ state is encountered at a variable latch along the beam path, the latch switches the beam path between the upper and lower branches.  If the output power of the final latch is in the upper (lower) branch, then the parity of the variable assignment is odd (even), and the SET (RESET) port of the check latch receives power, driving the check latch into the $|\text{unsatisfied}\rangle = \one$ ($|\text{satisfied}\rangle = \zero$) state.  The rate at which the check latch is driven to the appropriate state is proportional to the input field power $|\alpha|^2$ in units of photons per second.

The check latch $Q^{\text{check}}_\mathbf{c}$ in turn routes fields that participate in the feedback circuit described in the next Section.

%%%%%%%%%%%%%%%%%%%%%%%%%%%%%%%%%%%%%%
\subsubsection{Feedback to variables} \label{sec:FeedbackToVariables}

\begin{figure}%Feedback toVariables%%%%%%%%%%%%%%%%%%%%%%%%%%%%%%%
\includegraphics[width=8.6cm]{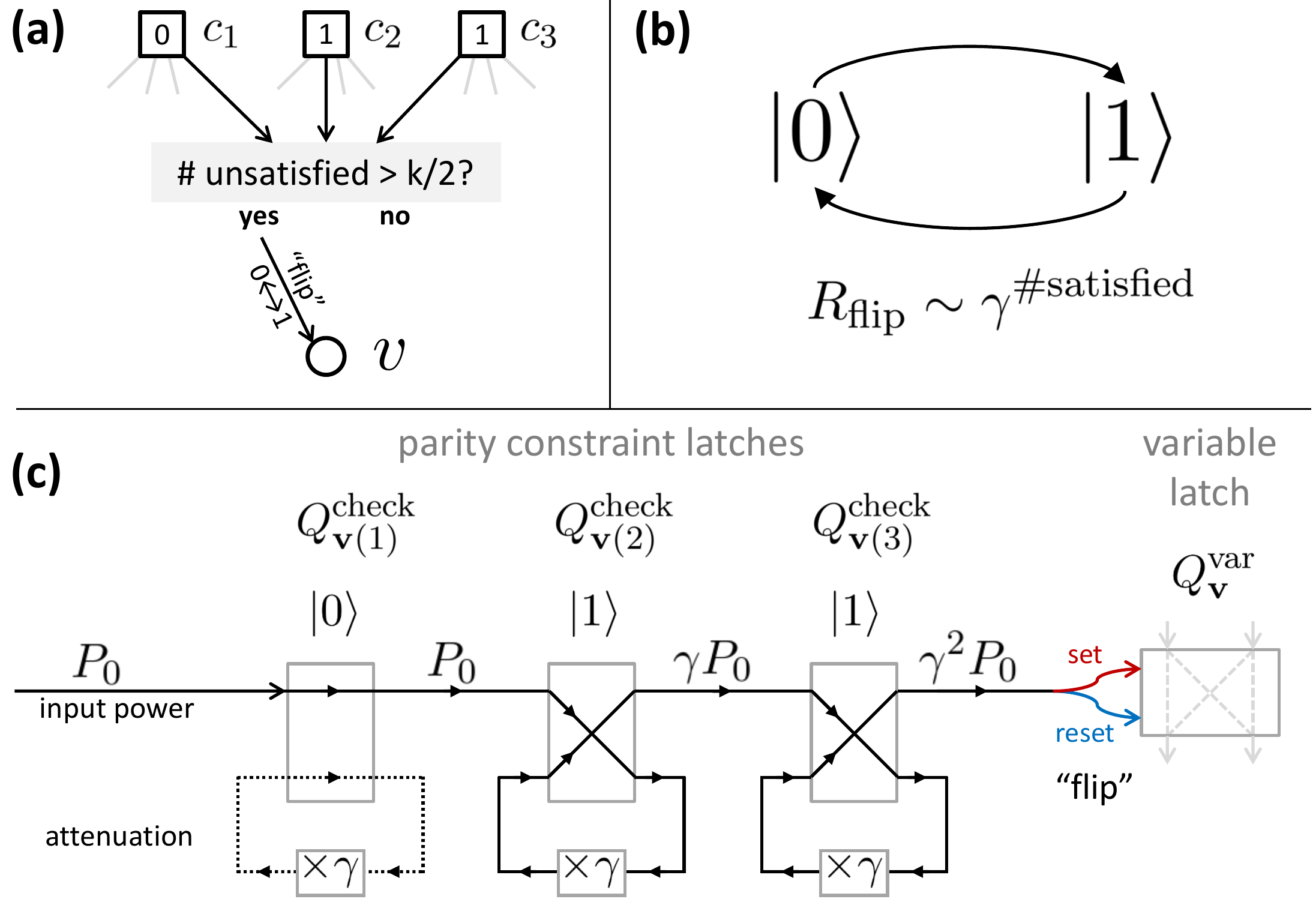}
\caption{Photonic circuit implementing feedback to variables.  \textbf{(a)}  The ``flip" operation for a fragment of the Tanner graph of a linear error-correcting code (rest of graph in gray).  If a majority of the parity check constraints that include variable $v$ are unsatisfied, the assignment of variable $v$ is flipped \cite{SipserSpielman1996}.  \textbf{(b)} For our photonic circuit implementation, the rate of ``flipping" the variable system state scales exponentially with the number of satisfied parity check constraints.  \textbf{(c)} The photonic circuit implementation of the error-correcting feedback using check latch systems $Q^{\text{check}}_{\mathbf{v}(1)},\ldots,Q^{\text{check}}_{\mathbf{v}(l)}$ and variable latch system $Q^{\text{var}}_\mathbf{v}$.  Power driving the variable system $Q^\text{var}_\mathbf{v}$ to flip is attenuated by a factor of $\gamma$ for every satisfied parity check constraint.}\label{fig:FeedbackToVariables}
\end{figure}

Fig. \ref{fig:FeedbackToVariables} shows our feedback to variables construction.  For a variable $v$, let  $\mathbf{v}$ denote the $l$ parity check constraints that include $v$: $\mathbf{v} =  \{\mathbf{c} : v \in \mathbf{c}\}$.  The current value (parity - $0$ or $1$) of the each check in $\mathbf{v}$ is represented by the state ($\zero$ or $\one$) of latches $Q_{\mathbf{v}(1)}^{\text{check}},\ldots,Q_{\mathbf{v}(l)}^{\text{check}}$.  As shown in Fig. \ref{fig:FeedbackToVariables}(c), the check latches share a common optical path.  An input field with amplitude $\beta$ is incident to input port in$_1$ of latch $Q_{\mathbf{v}(1)}^{\text{check}}$.  Subsequently, for each check latch $Q_{\mathbf{v}(i)}^{\text{check}}$, $1\leq i \leq l$, the second output is fed back into the second input of the same latch after passing through an attenuator (e.g. a beamsplitter) that dumps (e.g. reflects out of the beam path) a fraction $\gamma < 1$ of incident power and transmits a fraction $1-\gamma$ of the power back into the beam path.

%%%%%%%%%%%%%%%%%%%%%%%%%%%%%%%%

Each time an unsatisfied parity check constraint state ($\one$ state) is encountered at a check latch along the beam path, the power reaching the next check latch in the path is attenuated by a factor of $\gamma$.  The output of the final check latch in the path $Q_{\mathbf{v}(l)}^{\text{check}}$ is routed to drive both the SET and RESET inputs of the variable latch $Q^{\text{var}}_v$, causing it to ``flip" between the $\zero$ and $\one$ states.

Once a flip of variable $v$ occurs, the parity check system discussed in the previous Section updates the states of the check systems that include this variable, resulting in an updated value of the flipping rate for variable $v$.  If the power in the measurement circuit used to perform the parity check computation is low enough, the feedback circuit may induce multiple flips of the same variable before the measurement system reacts.  We consider this situation in the numerical results Section below.

The rate at which the variable latch $Q^{\text{var}}_v$ flips is proportional to the attenuated power outgoing from the final latch in the beam path:

\beq
R_\text{flip} \sim  \gamma^{(l - \text{\#unsat. checks})}\ |\beta|^2 = \gamma^\text{\#sat. checks}\ |\beta|^2
\label{eq:RateFlip}
\eeq

%\subsubsection{Feedback-induced errors}

If all $l$ parity constraints that include a variable $v$ are unsatisfied, the state of variable latch $Q^{\text{var}}$ flips with the maximum rate proportional to $|\beta|^2$.  If all $l$ constraints are satisfied, the variable is flipped with non-zero rate proportional to $\gamma^l |\beta|^2$.  Thus our circuit can induce errors.  For $\gamma \ll 1$, a single induced error should be quickly corrected since the rate for correcting it is a factor of $1/\gamma^l \gg 1$ larger than the rate for inducing it.

Our circuit corrects errors that are involved in $i$ parity check violations on a timescale proportional to $1/\gamma^i$.  The smaller we make the attenuation factor $\gamma$, the fewer induced errors there are, but the longer the decoding takes to complete.  We derive some bounds on the maximum value of $\gamma$ in terms of the code parameters such that our procedure is likely to succeed in Appendix \ref{app:FeedbackStrengthGammaBounds}.  We guess that the attenuation factor $\gamma$ should not be too small, since the decoding probability may increase when some induced errors are permitted, as observed in \cite{SipserSpielman1996}.  This intuition is consistent with our observations in the numerical results Section below.

\begin{figure}%ParityCheckFeedbackCircuitFragment%%%%%%%%%%%%%%%%%%%%%%%%%%%%%%%%
\includegraphics[width=8.6cm]{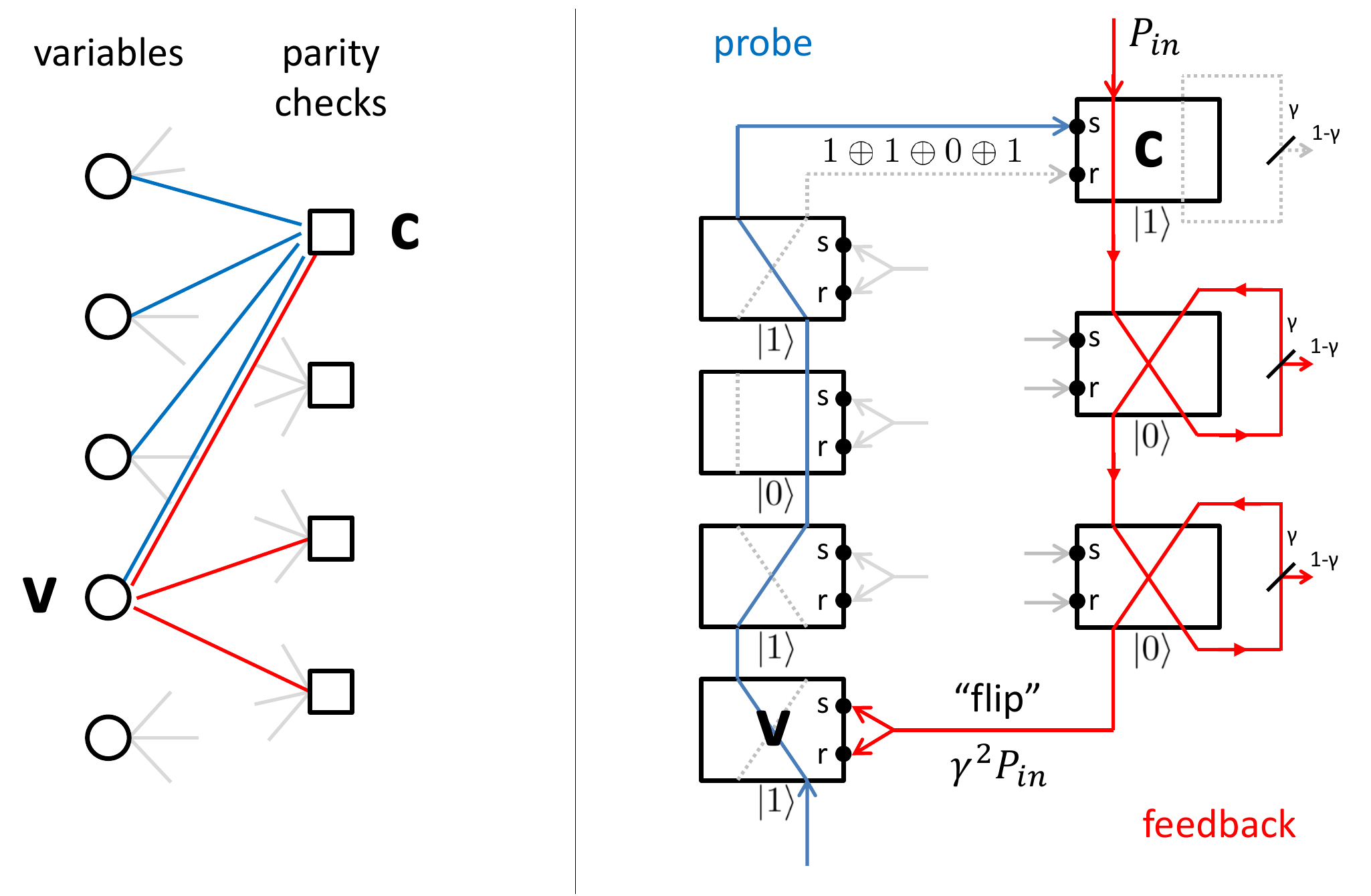}
\caption{Measurement and feedback circuit fragment (right) corresponding to fragment of Tanner graph (left).}\label{fig:ParityCheckFeedbackCircuitFragment}
\end{figure}%%%%%%%%%%%%%%%%%%%%%%%%%%%%%%%%

\begin{figure}%ParityCheckFeedbackCircuitFragment%%%%%%%%%%%%%%%%%%%%%%%%%%%%%%%%
\includegraphics[width=8cm]{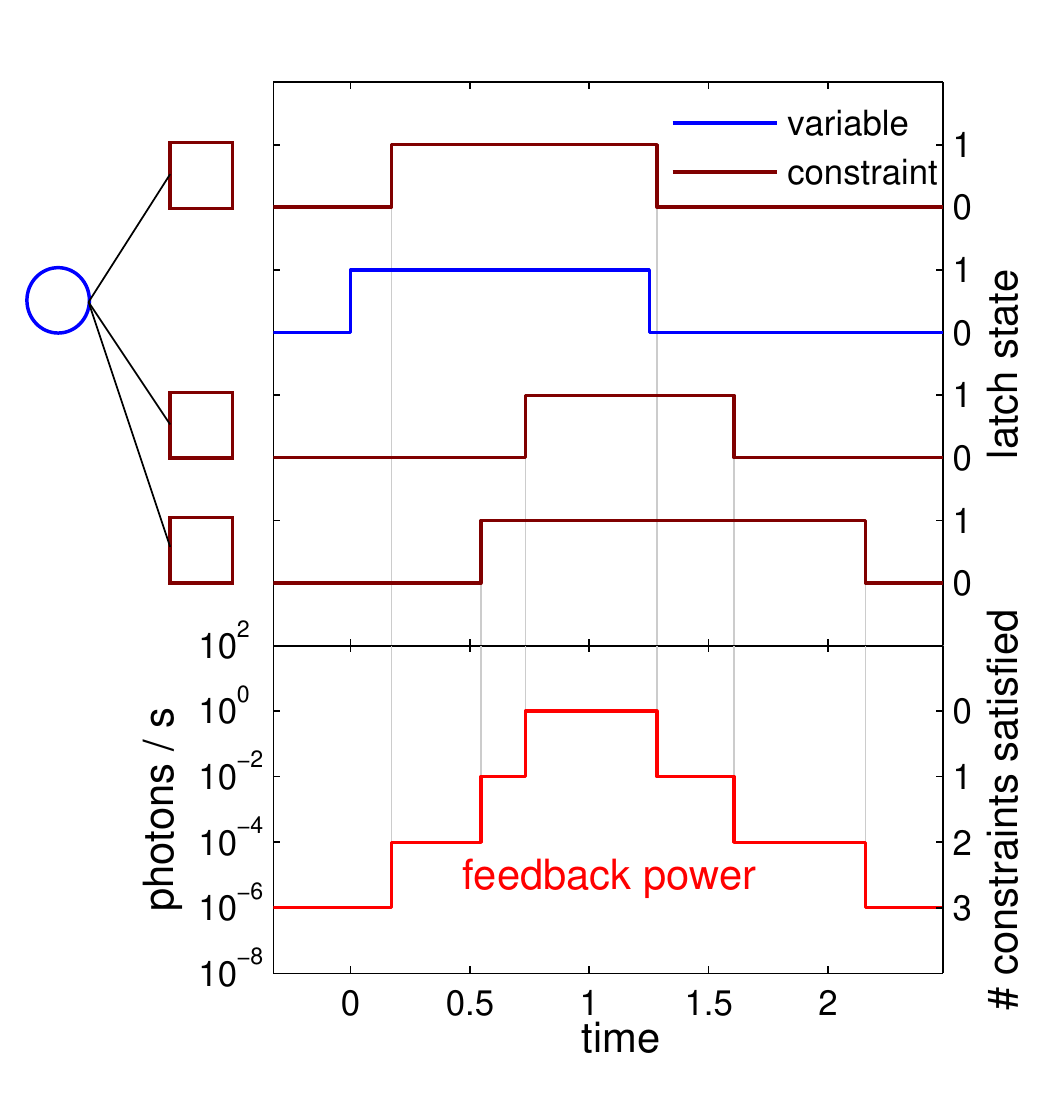}
\caption{Part of a trajectory of the decoding circuit for a fragment of an error-correcting code.  (top panel, blue line) state of a latch corresponding to a variable bit.  (top panel, dark red lines) states of latches corresponding to parity check constraints that include the blue variable bit.  (bottom panel) the feedback power applied to the variable bit, inducing it to ``flip" state.  On a log scale, this feedback power is proportional to the number of satisfied parity check constraints that include this variable bit.  See text for trajectory narration.}\label{fig:TrajectoryFragment}
\end{figure}%%%%%%%%%%%%%%%%%%%%%%%%%%%%%%%%

\subsection{Complete circuit summary plots}

Fig.~\ref{fig:ParityCheckFeedbackCircuitFragment} shows both the measurement and feedback subcircuits for a fragment of our decoder circuit corresponding to a fragment of the Tanner graph of an error-correcting code.  There is one such fragment for each of $n l$ edges in the Tanner graph of the code.

Fig.~\ref{fig:TrajectoryFragment} shows a portion of a simulated trajectory for a fragment of the code.  The top panel shows the state ($\zero$ or $\one$) of a latch corresponding to a variable bit (blue) and the three latches corresponding to the three parity checks that include this bit (dark red).  At time $0$, an error causes the variable bit latch (blue) to flip state (perhaps the component malfunctioned or the feedback system induced the error).  The three check latches corresponding to this bit then turn on (enter the unsatisfied, $\one$ state) after some exponentially-distributed waiting time (the mean of the waiting time is set by the input probe power used to perform the parity check sum computation).  For each check latch that enters the unsatisfied $\one$ state, the feedback power reaching the variable bit grows by a factor of $1/\gamma$, where $\gamma$ is the attenuation constant.  Around time $1.25$, the feedback induces the bit to flip back to the $\one$ state.  After an additional random waiting time, the three latch systems return to the satisfied $\zero$ state.  Note that the feedback power reaching the bit is never $0$, but reaches a minimum when all parity check constraints are satisfied.

%%%%%%%%%%%%%%%%%%%%%%%%%%%%%%%%%%%%%%
%\subsubsection{Initialization}

%[include this section?] We must load the channel output into the initial assignment of the variable systems.  This can be done by driving the set or reset port of the variable systems conditional on the state of the channel [assume transducer?]

\subsection{Fan-in/Fan-out}

Our decoder circuit requires each variable latch component to participate in multiple ($l$) parity check constraints, and requires each parity constraint latch component to feed back to multiple ($k$) variables.  Since the latch described in Section \ref{sec:Components} (and in greater detail in Appendix \ref{app:Latch}) can switch only a single pair of signal inputs, it is not on its own sufficient for our needs.  We can augment our latch to achieve the desired fan-in/fan-out (and avoid the difficulty of having multiple beam paths access a single structure in a planar circuit) by breaking up each latch into a set of subsystems, each responsible for routing a single in/out signal pair. The subsystems are yet more latches, but correspond to the single pair of in/out signals latch description of Section~\ref{sec:Components}.  This augmented latch is used implicitly in our circuit description above and is described in Appendix \ref{app:faninfanout}.

%%%%%%%%%%%%%%%%%%%%%%%%%%%%%%%%%%%%%%
\section{Numerical Experiments}
%
%We show several trajectories of our decoder's time evolution, plot several measures of performance - successful decoding probability, mean decoding time, mean decoding energy - versus parameters in our construction, and present a direct numerical comparison to the results of \cite{SipserSpielman1996}.

Table~\ref{table:parameters} lists the parameters used in our simulations. The spontaneous flip rate $\eta$ models component noise during the computation---all latches in our circuit independently flip ($\zero \leftrightarrow \one$) with rate $\eta$.

\begin{table}[ht]
\centering
\begin{tabular}{c|c|c}
\hline \hline
parameter & symbol & notes \\ \hline
block length & $n$ & \\
checks per variable & $l$ & $m = nl/k$ parity checks\\
variables per check & $k$ & \\ \hline
\begin{tabular}{c}probe intensity \\ feedback intensity \\ \end{tabular} & \begin{tabular}{c} $\alpha_\text{pr}$ \\ $\alpha_\text{fb}$ \\ \end{tabular}  & power $\sim|\alpha|^2$ \\ \hline
\begin{tabular}{c}feedback power \\ attenuation \\ \end{tabular} & $0 < \gamma < 1$ & \begin{tabular}{c} rate to flip variable\\ $= |\alpha_\text{fb}|^2 \gamma^\text{\#satisfied checks}$  \\ \end{tabular}  \\ \hline
spontaneous flip rate & $\eta$ & \begin{tabular}{c}all latches independently \\ flip state with rate $\eta$ \\ \end{tabular} \\
\hline \hline
\end{tabular}
\caption{Simulation parameters}
\label{table:parameters}
\end{table}

%%%%%%%%%%%%%%%%%%%%%%%%%%%%%%%%%%%%%%
%\subsection{A note on quantum trajectories [possibly should be an appendix]}
\subsection{Simulating quantum trajectories} \label{sec:trajectories}

Our circuit evolves according to the master equation (\ref{eq:mastereq}). Rather than solve this equation for the density matrix $\rho$ for our system, we sample multiple trajectories of the system wavefunction $|\psi\rangle$ and average observed quantities over these trajectories. Simulation of quantum trajectories given a master equation in the form of (\ref{eq:mastereq}) is computationally easier than integrating the master equation and is discussed in detail in \cite{Wiseman1996}. One way to perform such simulations is to sample exponentially-distributed jump times for each component of the system $\mathbf{L}$ vector (rate for $i$-th component is $\sim |\langle \psi|L_i^\dagger L_i |\psi\rangle|^2$), apply the nearest-in-time jump to the system wavefunction, and resample all of the jump times given the new wavefunction. In general, there is a smooth Hamiltonian evolution occurring between jumps as well, but our decoder circuit's Hamiltonian is diagonal in the $\{\zero, \one\}$ state basis, and this basis is fixed by the components of $\mathbf{L}$ (the jump terms) so we can ignore the smooth evolution and treat the system as a continuous time Markov jump process.

We prefer the trajectory approach in part because we want to average over different random instances of the expander code (with different network connectivities each time) and because it is useful to examine the time evolution of individual trajectories for an intuitive view of the circuit.

%%%%%%%%%%%%%%%%%%%%%%%%%%%%%%%%%%%%%%
\subsection{Trajectories}

\begin{figure}%Trajectories Decode%%%%%%%%%%%%%%%%%%%%%%%%%%%%%%%
\includegraphics[width=8.6cm]{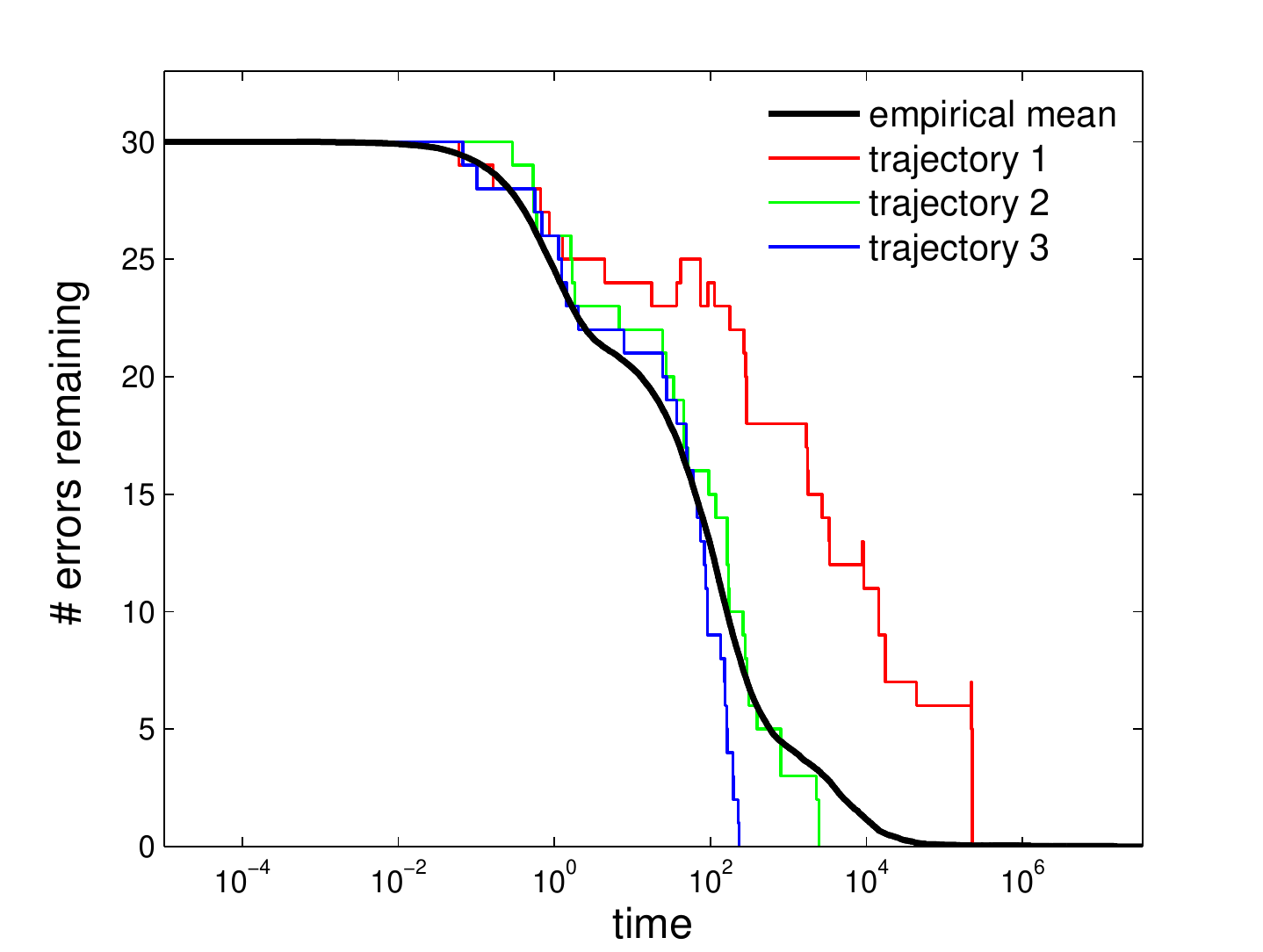}
\caption{Trajectory simulation of the iterative photonic decoder for a $(n = 1000, l = 5, k = 10)$ expander code, $30$ initial errors, circuit parameters $\gamma = 0.01$, $|\alpha_\text{fb}|^2 = 1$, $|\alpha_\text{probe}|^2 = 10^5$, $\eta=0$.  (black) the mean number of errors remaining vs. time averaged over 999 trajectories.  (red, green, blue) the number of errors remaining vs. time for three individual trajectories.}\label{fig:TrajectoriesDecode}
\end{figure}

We uniformly randomly sample 30 bits to corrupt from the initial all-0 codeword of length $n = 1000$ for a randomly sampled LDPC code with $l=5$, $k=10$, and track the remaining number of errors in time.  The code is generated by randomly sampling a bipartite graph with 1000 variable nodes each with degree 5, 500 check nodes each with degree 10.  We take the feedback attenuation parameter $\gamma = 0.01$, set the feedback power to 1 (arbitrary units), the probe power to something much larger ($10^5$), and set the rate for spontaneous component flips $\eta=0$.  Fig. \ref{fig:TrajectoriesDecode} shows the number of errors remaining as a function of time averaged over 999 trajectories, and for three individual trajectories.  999 of 1000 trajectories decoded successfully (converged the all-0 codeword).  The one that did not is not included in the average.  %Note that we show trajectory time - the sum of the waiting times between simulated state jumps - rather than CPU time to run the simulation.

%%%%%%%%%%%%%%%%%%%%%%%%%%%%%%%%

We point out two features of the trajectory simulations.  One is that (e.g. the red trajectory in Fig. \ref{fig:TrajectoriesDecode}) the number of errors remaining sometimes increases in the course of a simulation.  As discussed in our circuit description in Section \ref{sec:FeedbackToVariables}, the circuit induces errors at some non-zero rate and then corrects the induced errors.  Errors are most likely to be induced for variables that are involved in some, but not a majority of parity check violations.  When the attenuation constant $\gamma$ is too high (too little attenuation), the circuit may induce errors faster than they are corrected, resulting in a failure to decode.  On the other hand, as $\gamma$ is decreased, the circuit corrects errors at a lower rate, suggesting an optimal value of $\gamma$ in terms of a performance vs. decoding time tradeoff.  This tradeoff is considered in the next Section.

Second, the empirical mean of 999 trajectories (black trace in Fig. \ref{fig:TrajectoriesDecode} exhibits three shoulders (alternates between being locally convex and concave) in its decay toward 0.  The shoulders are spaced approximately $1/\gamma = 100$ logarithmic time units apart, corresponding to the correction of errors that are involved in 5, 4, and 3 parity check violations, respectively.  The mean number of errors remaining first declines significantly at time $t \sim 10^0$, consistent with feedback at maximal rate (no attenuation) $|\alpha_\text{fb}|^2 = 1$ flipping variables all $l = 5$ of whose corresponding parity check constraints are initially unsatisfied.

%%%%%%%%%%%%%%%%%%%%%%%%%%%%%%%%%%%%%%
\subsection{Performance vs. initial number of errors}
%\subsection{Probability to decode vs. initial number of errors}

\begin{figure}%Trajectories Decode compare to SS96%%%%%%%%%%%%%%%%%%%%%%%%%%%%%%%
\includegraphics[width=9.2cm]{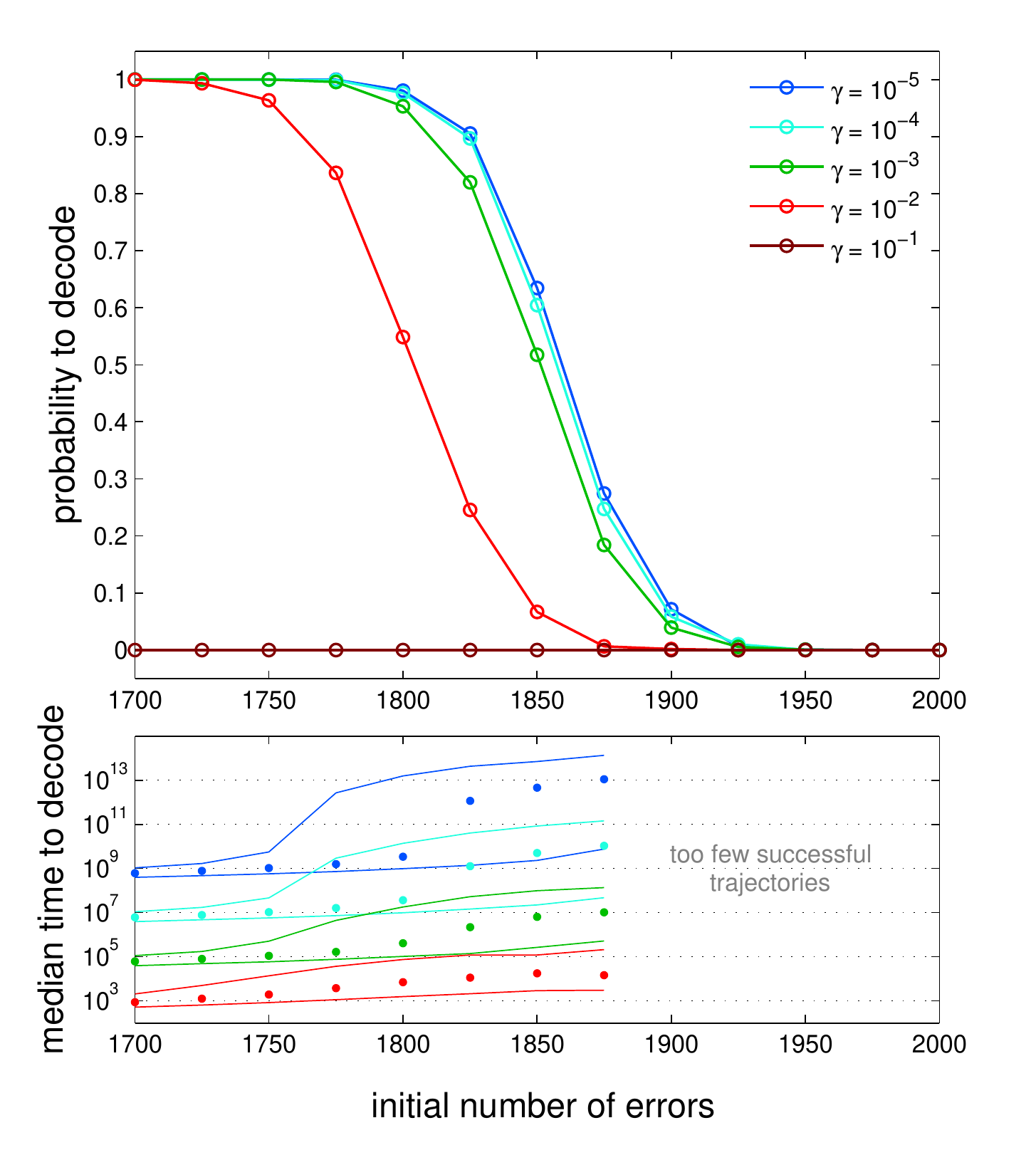}
\caption{Decoder performance with varying feedback power attenuation parameter $\gamma$.  (top) Probability to successfully decode vs. initial number of errors for several values of attenuation parameter $\gamma$ (see legend box).  (bottom, points) median time to decode conditioned on successfully decoding all errors.  (bottom, solid lines) 90\% interval for time to decode successfully.  We did not track these quantities past 1875 initial errors due to low succesful decoding probability.  The code parameters are the same as in \cite{SipserSpielman1996}: $(n = 40000,l=5,k=10)$.  We set $\alpha_\text{probe} = 10^3$, $\alpha_\text{fb} = 10$, $\eta = 10^{-80}$.  We sampled 3000 trajectories for each data point.}\label{fig:pdecode}
\end{figure}

We simulate our decoding circuit using the same code parameters as \cite{SipserSpielman1996}: a $(n = 40000,l=5,k=10)$ expander code, generated by randomly sampling a bipartite graph with 40000 variable nodes, 20000 check nodes, and degree 5 and 10 at the variable and check nodes, respectively.  The performance of our decoder in simulation for these parameters is shown in Fig. \ref{fig:pdecode}.  This performance (top panel) is somewhat better than that of \cite{SipserSpielman1996}'s scheme and somewhat worse than their version of the scheme permitting some backwards progress - occasionally allowing the total number of parity constraint violations to increase.

We see in Fig. \ref{fig:pdecode} (top) that the decoder's performance in terms of block error rate appears to saturate as the attenuation parameter $\gamma$ decreases.  At the same time, the median time~\footnote{We use the median, rather than the mean, time because as the probability to successfully decode drops sharply around 1800 initial errors, the distribution of decoding times spreads out over orders of magnitude (see Fig. \ref{fig:pdecode}, bottom), with the mean dominated by the few longest trajectories.  We imagine in actual use, we would operate some distance (in number of initial errors) below the point at which the decoder breaks down.} to successfully decode grows as $\gamma$ decreases (bottom), since the rate to flip bits scales exponentially in $\gamma$ (eq. (\ref{eq:RateFlip})).  Thus we could set $\gamma$ to the highest achievable value for a given channel error probability, desired mean decoding time, and probability to decode successfully.

%%%%%%%%%%%%%%%%%%%%%%%%%%%%%%%%

%%%%%%%%%%%%%%%%%%%%%%%%%%%%%%%%%%%%%%
\subsection{Performance vs. input power with noisy circuit components}

\begin{figure}%Performance vs. input power%%%%%%%%%%%%%%%%%%%%%%%%%%%%%%%
\includegraphics[width=8.6cm]{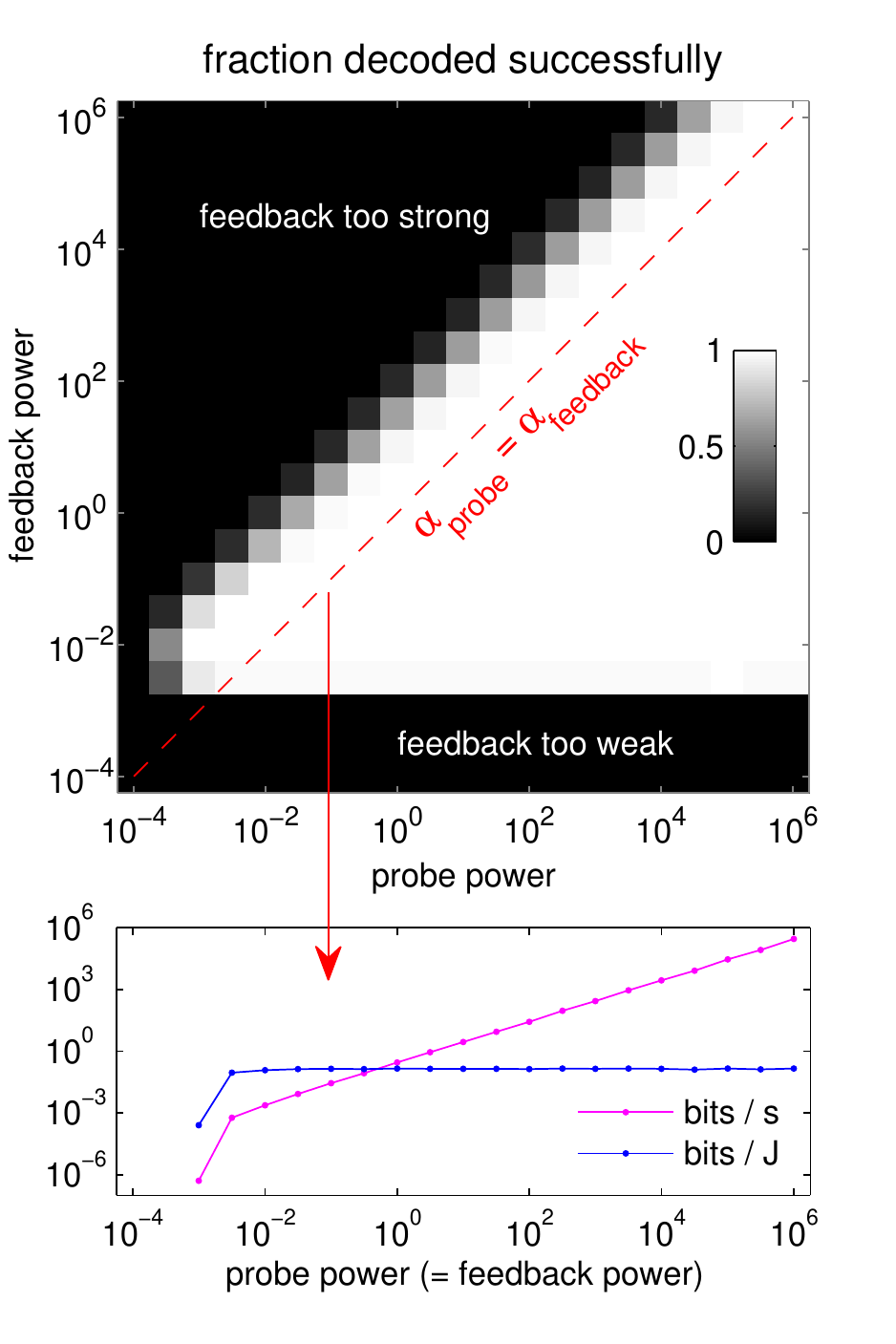}
\caption{Decoder performance with varying probe and feedback power in the presence of component noise.  (top, grayscale) Fraction of trials that decoded all of the initial errors succesfully vs. probe and feedback power.  (top, red dashed line) fixed probe to feedback power ratio (ratio value 1).  (bottom) Section of top plot (marked by red dashed line) corresponding to a fixed probe to feedbak power ratio.  (bottom, magenta) mean decoding rate in bits/unit time.  (bottom, blue) mean decoding energy (ratio of mean decoding rate and input power). Both performance measures are conditioned on successfully decoding all errors.  Component spontaneous flip rate $\eta = 10^{-8}$.  $\gamma = 0.01$.  Expander LDPC code parameters: block length $40000$, $l = 5$, $k = 10$, $1700$ initial errors.  We sampled 3000 trajectories per grayscale point.}\label{fig:VaryFeedbackProbePower}
\end{figure}

We consider the decoder's performance as a function of applied input power in terms of probability to decode, decoding rate (bits/s), and decoding energy (bits/J).  Additionally, we set some non-zero rate $\eta$ at which the circuit components undergo spontaneous flips ($\zero \leftrightarrow \one$).  This noise affects both the variable and the check latches and in turn both the measurement and feedback parts of the circuit.  Fig. \ref{fig:VaryFeedbackProbePower} shows our numerical results for fixed component noise rate $\eta$, LDPC code parameters, initial number of errors, and attenuation parameter $\gamma$ (see caption for parameter values).

We see (top panel of Fig. \ref{fig:VaryFeedbackProbePower}) that to decode successfully most of the time, the feedback power needs to be large enough to overcome the errors induced by noise in the circuit components, but not much larger than the probe power.  When the feedback power is much larger than the probe power, the probe circuit is too slow to turn off the feedback once an error is corrected and too slow to turn on the feedback for new errors (induced by either the feedback or spontaneous flips), so the feedback system may induce more errors than it corrrects.

For the bottom panel of Fig. \ref{fig:VaryFeedbackProbePower} we fixed the probe to feedback power ratio at 1 and plotted the mean decoding rate and energy versus input power in bit/s, bit/J, respectively~\footnote{These are in arbitrary time and energy units proportional to s and J, since we did not give physical values for our simulation parameters.  To compute a fiducial time and energy to decode, we reference the latch switching time estimated in~\cite{Mabuchi2009} using the parameters of \cite{Beau09} for a gallium phosphide photonic resonator and diamond nitrogen-vacancy system: $\tau_\text{sw} \approx 7\mu$s per switch at 1pW set/reset input power.  For 1700 initial errors in Fig. \ref{fig:pdecode}, we see (bottom panel) the time to decode scales as $1/\gamma^2$, suggesting that the time to correct errors that satisfy 2 out of 5 parity check constraints dominates the total decoding time.  The feedback power to correct these errors is attenuated by a factor of $\gamma^2$, so the mean time for switching them is $\tau_\text{sw} / \gamma^2$.  Setting the feedback power to 1pW, $\gamma = 10^{-2}$ (this is the highest value of $\gamma$ shown in Fig. \ref{fig:pdecode} such that the decoder succeeds for most trajectories with 1700 initial errors), we estimate the time to decode to be $\sim \tau_\text{sw}/\gamma^2 = 7\mu\text{s} / (10^{-2})^2 = 70\text{ms}$, the energy to decode to be $(\text{time to decode})\cdot(\text{input power}) = 70\text{ms}\cdot 1\text{pW} = 70\text{fJ}$ per latch;  The set/reset inputs of each latch in the circuit receive 1pW input power, so this estimate gives the energy to decode per latch.  There are 60,000 total latches in the decoder circuit for the parameters in Fig. \ref{fig:pdecode} (40,000 variables, 20,000 checks), so the total energy to decode is $\sim (\text{energy per latch})\cdot(\text{number of latches}) \approx 70\text{fJ}\cdot 60,000 \approx 4\text{nJ}$.}.  We defined the decoding rate as the reciprocal of the mean decoding time, conditioned on successfully decoding, and the decoding power as the decoding rate divided by the input power.

We see that for large enough input power, the decoding rate is proportional to the input power, while the energy cost per decoded bit is constant.

%%%%%%%%%%%%%%%%%%%%%%%%%%%%%%%%

%%%%%%%%%%%%%%%%%%%%%%%%%%%%%%%%%%%%%%
\section{Discussion}

%[Discussion?]

We have described a photonic circuit that implements an iterative decoding scheme for expander LDPC codes.  This circuit consists of a collection of optical latching relays, whose interactions via coherent fields map naturally onto the subroutines of the iterative decoder.

This circuit is autonomous---it is powered by the same optical signals that it acts upon to implement the decoding procedure, and it requires no external controller, measurement system, or clock signal. It operates robustly in the low-power limit in which quantum fluctuations of the optical fields are significant.  The feedback-induced latch state fluctuations provide a natural source of randomness to drive the decoding algorithm.  Crucially for the feasibility of such a system, our circuit's performance, as measured by decoding time and error rate, can be tuned smoothly by varying the optical input power. Tuning the input power can be done without loss in efficiency, as our circuit decodes a constant number of bits per Joule at a rate linear in the input power. Thus, noise that acts on the circuit components and potentially disrupts the computation can be overcome by increasing input power until the circuit works.

Our construction highlights the computational utility of cavity QED-based nanophotonic components for ultra-low power classical information processing, and points to the utility of the probabilistic graphical model framework in engineering autonomous optical systems that operate robustly in the quantum noise regime.

\begin{acknowledgements}
This work has been supported by the ARO (W911NF-13-1-0064) and DARPA (N66001-11-1-4106). DSP acknowledges the support of a Stanford Graduate Fellowship.
\end{acknowledgements}

%%%%%%%%%%%%%%%%%%%%%%%%%%%%%%%%%%%%%%
\appendix

%%%%%%%%%%%%%%%%%%%%%%%%%%%%%%%%%%%%%%
\section{Gough-James circuit algebra}\label{app:GoughJamesCircuitAlgebra}

We briefly review the Gough-James treatment of open quantum systems and circuits composed of such systems \cite{GoughJames2008,GoughJames2009}.  We give sufficient detail for the reader to reproduce our numerical simulations.

%%%%%%%%%%%%%%%%%%%%%%%%%%%%%%%%%%%%%%
\subsection{Open quantum systems}\label{app:OpenQuantumSystems}

In the Gough-James circuit algebra for modeling open quantum systems, a system coupled to $n$ external fields is parametrized by a $(\mathbf{S}, \mathbf{L}, H)$ triplet, where the scattering matrix $\mathbf{S}$ is $n$ by $n$ unitary with operator-valued entries, the coupling vector $\mathbf{L}$ is $n$ by 1 with operator-valued entries, and $H$ is the system's Hamiltonian.  Fig. \ref{fig:SLH} summarizes this picture.  The density matrix $\rho$ for the system's internal degrees of freedom evolves in time according to the master equation (eq. (\ref{eq:mastereq})):

\begin{figure}[]%SLH%%%%%%%%%%%%%%%%%%%%%%%%%%%%%%%%
\begin{center}
\includegraphics[width=8.6cm]{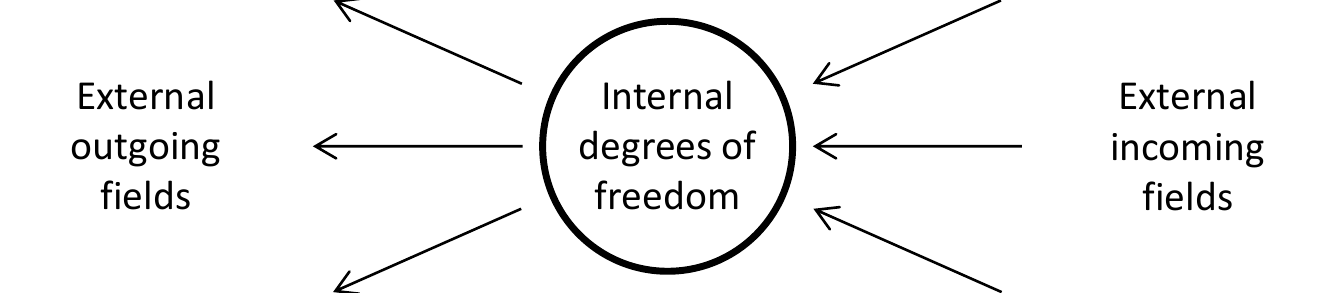}
\caption{A quantum system with internal degrees of freedom coupled to external fields.}\label{fig:SLH}
\end{center}
\label{eq:mastereqinapp}
\end{figure}%%%%%%%%%%%%%%%%%%%%%%%%%%%%%%%%

\beq
\dot{\rho}_t = -i  [H,\rho_t] + \sum_{i=1}^n \left(L_i \rho_t L_i^\dagger - \frac{1}{2}\{L_i^\dagger L_i ,\rho_t\}\right)
\eeq

\noindent where $[A,B] = AB-BA$, $\{A,B\} = AB+BA$, and $\dagger$ denotes conjugation. The scattering matrix $\mathbf{S}$ does not appear in (\ref{eq:mastereq}), but appears when we interconnect such systems below.

%%%%%%%%%%%%%%%%%%%%%%%%%%%%%%%%%%%%%%
\subsection{Circuits}\label{app:Circuits}

The Gough-James circuit algebra allows us to compute new $\SLH$ triplets in terms of old for two systems connected in series, in parallel, or for one system self-connected through feedback.  We briefly state these circuit composition rules.

The series product takes two open quantum systems $G_1 = (\mathbf{S}_1,\mathbf{L}_1,H_1)$, $G_1 = (\mathbf{S}_2,\mathbf{L}_2,H_2)$ coupled to an equal number of external modes and returns the system $G_2 \lhd G_1$ obtained by feeding the outputs of $G_1$ into the inputs of $G_2$:

\beq
G_2 \lhd G_1 = \left(\mathbf{S}_2 \mathbf{S}_1,\ \mathbf{S}_2 \mathbf{L}_1 + \mathbf{L}_2,\ H_1 + H_2 + \Im\left(\mathbf{L}_2^\dag \mathbf{S}_2 \mathbf{L}_1\right)\right)
\eeq

The concatenation product takes two open quantum systems $G_1$ and $G_2$, coupled to $n_1$ and $n_2$ modes, respectively, and returns the system $G_2 \boxplus G_1$ obtained by considering the two systems as one system coupled to $n_1 + n_2$ modes and introducing no interactions between them:

\beq
G_1 \boxplus G_2 = \left(\left(\begin{array}{cc} \mathbf{S}_2 & \mathbf{0} \\ \mathbf{0} & \mathbf{S}_1 \end{array}\right), \left(\begin{array}{c}\mathbf{L}_1 \\ \mathbf{L}_2 \end{array}\right), H_1 + H_2\right)
\eeq

The feedback product takes a single open quantum system coupled to $n$ modes and returns the system $[G]_{k\rightarrow l}$ obtained by feeding back the $k$-th output mode to the $l$-th input mode, coupled to $n-1$ external modes.  The form of this product is given in \cite{GoughJames2008} (Section 5) (and in the notation used here in \cite{QHDL}, Appendix A).

%[Give feedback operation?]

%%%%%%%%%%%%%%%%%%%%%%%%%%%%%%%%%%%%%%
\section{Components}\label{app:Components}

We describe the components we need for our decoder circuit in terms of a $\SLH$ triplet, focusing on an intuitive input-output picture.

%%%%%%%%%%%%%%%%%%%%%%%%%%%%%%%%%%%%%%
\subsection{Beamsplitter}

To give an intuition for these systems and to specify a components we need, we first describe the beamsplitter as an open Markov quantum system.  A 50/50 beamsplitter has two input and two output ports and is parametrized by:
\beq
B = \left(\mathbf{S} =
\frac{1}{\sqrt{2}}\left(\begin{array}{cc}
1 & 1 \\
-1 & 1 \\
\end{array}\right),\ \ \
\mathbf{L} =
\left(\begin{array}{c}
0 \\
0 \\
\end{array}\right),\ \ \
H = 0\right)
\eeq

By examining the scattering matrix, we see that for a field incident into input port 1, half the power is transmitted into output port 1 and half is reflected into output port 2 with a $\pi$ phase shift.  The beamsplitter has no internal degrees of freedom that concern us here, so $\mathbf{L} = \mathbf{0}$ and $H = 0$.  The scattering matrix for a beamsplitter that transmits a fraction $\gamma < 1$ of incident power - our attenuation component - is a 2 by 2 rotation matrix with angle $\arccos \sqrt{\gamma}$.

%%%%%%%%%%%%%%%%%%%%%%%%%%%%%%%%%%%%%%
\subsection{Coherent input field}

A coherent field input is modeled as a Weyl operator $W_{\vec{\alpha}}$, which displaces $n$ vacuum inputs into  coherent states $|\alpha_1\rangle,\ldots,|\alpha_n\rangle$ with amplitudes $\alpha_1,\ldots,\alpha_n$:

\beq
W_{\vec{\alpha}} = \left(\mathbf{S} = \mathbf{1}_{n\times n},\ \ \ \mathbf{L} = \left(\begin{array}{c}\alpha_1\\ \vdots \\ \alpha_n \\\end{array}\right),\ \ \ H = 0\right)
\eeq

For example, driving the beam splitter above with $|\alpha\rangle$ in the first input and $|\beta\rangle$ in the second input results in the series connection:

\beq
\begin{array}{l}
B \lhd W_{(\alpha,\beta)} = \\
\left(\mathbf{S} =
\frac{1}{\sqrt{2}}\left(\begin{array}{cc}
1 & 1 \\
-1 & 1 \\
\end{array}\right),\ \ \
\mathbf{L} =
\frac{1}{\sqrt{2}}\left(\begin{array}{c}
\alpha + \beta \\
-\alpha + \beta \\
\end{array}\right),\ \ \
H = 0\right)
\end{array}
\eeq

\noindent resulting in the mixing of the two inputs in the two outputs, as we expect.

%%%%%%%%%%%%%%%%%%%%%%%%%%%%%%%%%%%%%%
\subsection{Latch}\label{app:Latch}

In terms of an $(\mathbf{S},\mathbf{L},\mathbf{H})$ triplet, the latch is given by the concatenation (parallel product) of two systems: $Q_\text{set-reset}$ accepts the set and reset inputs and drives the latch into the $\zero$ or $\one$ state, and $Q_\text{in-out}$ routes the input fields $\text{in}_{1,2}$ into the output fields $\text{out}_{1,2}$.  We have $Q = Q_\text{set-reset} \boxplus Q_\text{in-out}$, where

\begin{align}
Q_\text{set-reset} = & \left(\mathbf{S}_\text{set-reset} =
\left(\begin{array}{cc}
\Pi_0 & -\sigma_{10} \\
-\sigma_{01} & \Pi_1 \\
\end{array}\right), \right. \\ \nonumber
&\ \ \ \mathbf{L} = \left.
\left(\begin{array}{c}
0 \\
0 \\
\end{array}\right),
H = 0\right)
\end{align}

\beq
Q_\text{in-out} = \left(\mathbf{S}_\text{in-out} =
\left(\begin{array}{cc}
\Pi_0 & -\Pi_1 \\
-\Pi_1 & \Pi_0 \\
\end{array}\right),
\mathbf{L} =
\left(\begin{array}{c}
0 \\
0 \\
\end{array}\right),
H = 0\right)
\eeq

\noindent where $\Pi_0 = |0\rangle\langle0|$ and $\Pi_1 = |1\rangle\langle1|$ are projection operators onto the $\zero$ and $\one$ states and $\sigma_{01} = |0\rangle\langle1|$, $\sigma_{10} = |1\rangle\langle0|$ switch $\zero$ and $\one$.  Conditional on the state of the latch, either $\mathbf{S}_\text{in-out} = \left(\begin{array}{cc}1 & 0 \\ 0 & 1\\\end{array}\right)$ or $\mathbf{S}_\text{in-out} = \left(\begin{array}{cc}0 & -1 \\ -1 & 0\\\end{array}\right)$, thus either switching or not switching the input fields.  This is the same latch model as that used in our earlier work \cite{BitFlipCode}.

A possible physical system that achieves this desired behavior is shown in Fig.~\ref{fig:Relay} and was first proposed in \cite{Mabuchi2009}.  The $\zero$ and $\one$ states are degenerate ground states of an atom in a cavity.  Set, reset, and input fields are resonant with transitions to one of two excited states ($|e\rangle$ and $|s\rangle$), from which the atom then decays back into one of the ground states.  In a regime of strong atom-cavity coupling, the limiting behavior of the switch system is obtained by using the QSDE limit theorem \cite{BoutenvanHandelSilberfarb2008} to adiabatically eliminate the excited state dynamics.  An alternate proposal for such a switch using a Kerr cavity is found in \cite{Mabuchi2011}.

%%%%%%%%%%%%%%%%%%%%%%%%%%%%%%%%%%%%%%
\section{Bounds on nonlinearity of feedback}\label{app:FeedbackStrengthGammaBounds}

Consider a $(n,l,k)$ LDPC code, and suppose there is only one variable $v$ that needs to be flipped to return to a codeword.  This variable participates in $l$ parity check constraints, all of which are violated, so it flips at some maximal rate $r$.  These $l$ parity check constraints together include at most $l(k-1)$ variables other than $v$ (at most because they may have some in common), each of which is involved in least one parity constraint violation, and so flips with rate at least $r \gamma^{l-1}$.  The total rate for erroneously flipping any variable other than $v$ is then

$$R_\text{err} = l (k-1) \ r \  \gamma^{l-1} $$

In order to flip $v$ before errors accumulate, we set $r > R_\text{err}$ and find

$$ \gamma < \left(\frac{1}{l(k-1)}\right)^{\frac{1}{l-1}}$$

In our numerical tests we have used $l=5$, $k=10$, yielding $\gamma < 0.38$.  Numerically we found that our decoder mostly fails to decode already for $\gamma=0.1$ (see Fig. \ref{fig:pdecode}), but this is an upper bound assuming only one total error.

\section{Fan-in/Fan-out} \label{app:faninfanout}

As discussed in Section \ref{sec:CircuitConstruction}, our circuit requires a latch component $Q^\text{var}_\textbf{v}$ corresponding to variable bit $\mathbf{v}$ to participate in multiple ($l$) parity check constraints, and a latch $Q^\text{check}_\textbf{c}$ corresponding to parity check $\mathbf{c}$ to feed back to multiple ($k$) variables.  Since the latch described in Section  \ref{sec:Components} and Appendix \ref{app:Components} routes only two input and two output ports (in/out$_{1,2}$), it is insufficient for our needs: we need a latch that routes multiple in/out$_{1,2}$ signal pairs - switching each pair if and only if the latch state is $\one$ (see upper panel of Fig. \ref{fig:faninfanout}).   We can augment our latch to achieve the desired fan-in/fan-out in two ways.

\subsection{Routing multiple signals}

One way is to simply add extra input/output ports to the latch system depicted in Fig. \ref{fig:Relay}: we could have input pairs in$_{1,2}^{(i)}$ for $i \in \{1,\ldots,N\}$ and the corresponding outputs (in addition to the two set/reset ports) for some integer $N$, all coupled to the same latch state.  This would be difficult to achieve in a nanophotonic system, if only due to constraints of geometry - it would be difficult to have multiple beam paths access a single structure in a planar circuit.

An alternate scheme is depicted in Fig. \ref{fig:faninfanout}.  The idea is to break up each latch into a set of $N$ subsystems $Q_\text{route}^{(1)},\ldots,Q_\text{route}^{(N)}$, each responsible for routing a single in/out signal pair, and a single subsystem $Q_\text{sr}$ responsible for accepting the set/reset inputs (see Fig. \ref{fig:faninfanout}).  Each of the $N+1$ subsystems is another latch, but one that routes only a single in/out signal pair and fits the description of Section \ref{sec:Components}.  The set/reset subsystem $Q_\text{sr}$ routes power (in orange path in Fig. \ref{fig:faninfanout}) to the set/reset ports of the $N$ routing subsystems $Q_\text{route}^{(i)}$, driving the state of each routing subsystem to match the state of the set/reset subsystem.  Thus the $N$ routing subsystems $Q_\text{route}^{(i)}$ all mirror the overall system state, defined as the state of the set/reset subsystem $Q_\text{sr}$.

This construction introduces a delay in distributing the state of the set/reset subsystem to the $N$ routing subsystems - due to both the waiting time for a routing subsystem to switch and to the time for a signal to propagate around a circuit (we do not model the latter source of delay for this circuit).  The construction also introduces extra circuit components that could be subject to noise (e.g. spontaneously changing their state).  We thus need to use high-enough input power (in orange path in Fig. \ref{fig:faninfanout}) to make this construction useful.

\subsection{Accepting multiple set/reset inputs}

We note that we can use a similar construction to make a latch system that accepts multiple set/reset inputs in addition to routing multiple in/out signal pairs; though such a system does not appear in our decoder circuit, it may be useful for other purposes.  When there are multiple set/reset input pairs for a device, these inputs lose their interpretation as ``set" and ``reset" for the elecronic latch.  We can instead associate each set/reset pair with an internal state and define an overall state as the sum modulo 2 of these internal states, so that changing any of the internal states changes the overall state.  This behavior could be useful if we are interested in having a circuit component with multiple ``flip" control inputs.

\begin{figure}[]%faninfanout%%%%%%%%%%%%%%%%%%%%%%%%%%%%%%%%
\begin{center}
\includegraphics[width=8.6cm]{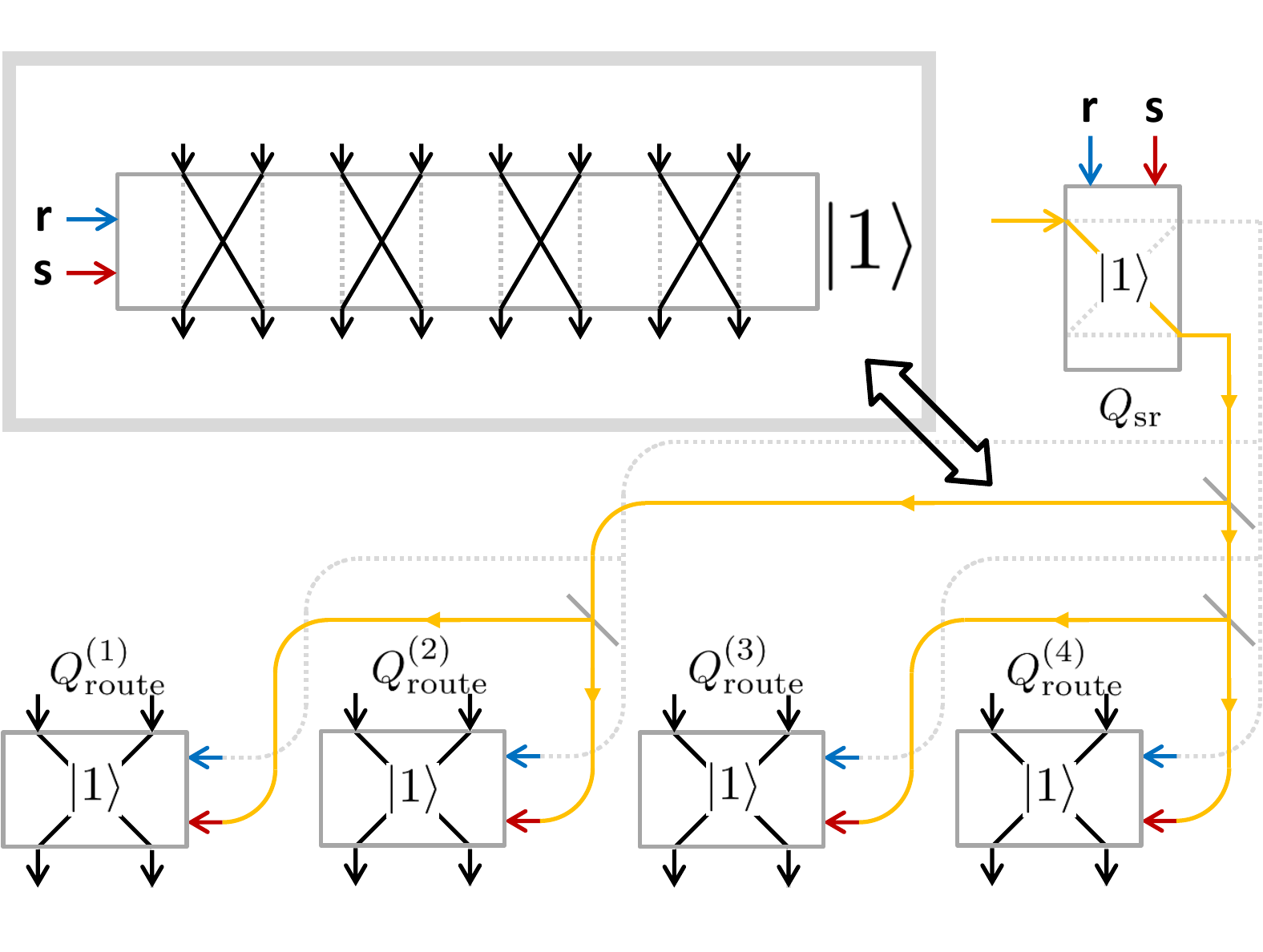}
\caption{(top left inset) A latch that routes multiple in/out signal pairs, as needed by our decoder circuit.  (main figure) implementation of latch in inset using only the single in/out signal pair latches described in Section \ref{sec:Components}.  The routing latches $Q_\text{route}^{(i)}$ are each responsible for routing a single in/out signal pair.  The set/reset latch $Q_\text{sr}$ accepts external set/reset inputs and is responsible for distributing its state - the overall latch state - to the routing latches.}\label{fig:faninfanout}
\end{center}
\end{figure}

The idea is to break up the set/reset latch subsystem described above into a set of $M$ subsystems $Q_\text{sr}^{(j)}$, $j \in \{1,\ldots,M\}$, each of which accepts only a single set of set/reset inputs.  The overall system state is then defined as the sum modulo 2 of the set/reset subsystems, so flipping the state of any of them changes the overall state.  The sum modulo 2 is performed as for the parity check circuit described in Section \ref{sec:ParityChecks} and shown in Fig. \ref{fig:ParityCheck} (b).  The probe beam path (black path in Fig. \ref{fig:ParityCheck} (b)) would now access each of the $Q_\text{sr}^{(i)}$ subsystems in sequence before driving the set or reset port of each of the $Q_\text{route}^{(j)}$ subsystems, as described in the previous Subsection (orange path in Fig. \ref{fig:faninfanout}).

%%%%%%%%%%%%%%%%%%%%%%%

\bibliography{draft_v2}

\end{document}